# Modelling the artificial night sky brightness at short distances from streetlights

Salvador Bará,[1,*]  Carmen Bao-Varela,[2] Miroslav Kocifaj[3,4]

[1] A. Astronómica 'Ío', 15005 A Coruña, Galicia.

[2] Photonics4Life Research Group, Facultade de Física e Facultade de Óptica e Optometría, Instituto de Materiais (iMATUS), Universidade de Santiago de Compostela, Campus Vida, E-15782 Santiago de Compostela, Spain.

[3] Faculty of Mathematics, Physics, and Informatics, Comenius University, Mlynská dolina, 842 48 Bratislava, Slovakia.

[4] ICA, Slovak Academy of Sciences, Dúbravská cesta 9, 845 03 Bratislava, Slovakia.

(*) email: salva.bara@usc.gal

**Abstract**

Contrary to some widespread intuitive belief, the night sky brightness perceived by the human eye or any other physical detector does not come (exclusively) from high in the sky. The detected brightness is built up from the scattered radiance contributed by all elementary atmospheric volumes along the line of sight, starting from the very first millimeter from the eye cornea or the entrance aperture of the measuring instrument. In artificially lit environments, nearby light sources may be responsible for a large share of the total perceived sky radiance. We present in this paper a quantitative analytical model for the sky radiance in the vicinity of outdoor light sources, free from singularities at the origin, which provides useful insights for the correct design or urban dark sky places. It is found that the artificial zenith sky brightness produced by a small ground-level source detected by a ground-level observer at short distances (from the typical dimension of the source up to several hundred meters) decays with the inverse of the distance to the source. This amounts to a reduction of 2.5 mag/arcsec$^2$ in sky brightness for every $\log_{10}$ unit increase of the distance. The effects of screening by obstacles are also discussed.

# 1. Introduction

The propagation of the artificial radiance in the nocturnal environment is a thriving field of research, due to its relevance for quantitative assessment of light pollution effects. A large effort has been devoted in the last decades to develop efficient conceptual and computational tools for modelling the artificial night sky brightness in large areas of the planet under different radiant and atmospheric conditions [1-5], helped by the progressive availability of high-quality data on the nighttime light sources obtained from low Earth orbiting platforms [6-16].

A growing degree of attention is also being paid to the spatial distribution of individual radiant sources (streetlights, commercial screen displays, and other) at sub-pixel scale, and its relationship with the mid-resolution, pixel-averaged information provided by satellite radiometers. Some examples of this line of work are the studies on the point-spread functions for individual luminaires [17], and the development of technologies and citizen science campaigns with capabilities for performing detailed wide-area light source inventories [18-26].

In this paper we address an open issue in this field: How large is the contribution of the nearby light sources to the total brightness of the night sky? Disentangling this contribution from the baseline, structural brightness produced by the myriad of light sources of districts and cities located around the observer is an instrumental step for devising public policies aiming to improve the quality of the local sky in artificially lit environments. Both contributions to the deterioration of the night sky must be addressed in parallel; however, the actions to undertake may be different in each case. Whereas the disruptive effects of nearby light sources (those located up to few hundred meters distance from the observer) can be controlled by means of local actions as e.g., effective screening, the management of the contributions of the surrounding territory, up to tens or hundreds of kilometers distance, requires an adequate territorial planning approach [27-30].

Two main reasons suggest that the relevance of nearby light sources may be far from negligible. One is the anecdotal evidence that the number of stars visible in the urban night sky may vary sharply if the observer moves a few tens or hundred meters within the city. An increase of one unit in the naked-eye star limiting magnitude is not exceptional in this context. The other is that the radiance from close light sources reaches the lower segments of the observer line of sight with little to no atmospheric attenuation and gets efficiently scattered due to the large density of the ground-level atmosphere. Let us recall that, contrary to some widespread intuitive belief, the night sky brightness perceived by the human eye or any other physical detector does not come exclusively (and sometimes even not mostly) from "high in the sky". The detected brightness is built up from the scattered radiance contributed by all elementary atmospheric volumes along the line of sight, starting from the very first millimeter from the eye cornea or the entrance of the detector. The local structure of the radiant field is then expected to be a relevant factor when it comes to evaluating the visual quality of the night sky.

For quantifying the propagation of artificial light at close distances from the sources we adopted here a single-scattering analytical model based on the one developed by Kocifaj in 2007 [4]. Care shall be exercised in the numerical integrations in order to avoid potential singularities at the origin. These singularities can be removed by using an exact expression for the solid angle subtended by an elementary source at all distances from the observer, including the limit for distance tending to zero. The use of a single-scattering model is justified in view of the short distances involved, under the assumption of not extremely large aerosol optical depths. The basic model is described in section 2. Besides the rigorous formulation developed in this section, some approximations are also derived for a simplified canonical case which provides useful physical insights about the dependence of the effects on the distance to the source. General numerical results deduced from the rigorous model are described in section 3. Discussion and conclusions are drawn in sections 4 and 5, respectively.

## 2. Methods

*2.1 Geometry of observation*

The basic configuration of the source-observer geometry is shown in Figure 1. A small light source at height $h_0 \geq 0$ above ground is located at an horizontal distance $D$ from an observer who, at a height $h_d \geq 0$, detects the artificial night sky brightness in any arbitrary direction of observation. This direction is specified by the unit vector $\boldsymbol{\alpha}(z,\varphi)$, where $z$ is the zenith angle and $\varphi$ is the azimuth of the line of sight, both measured in the observer's reference frame. For simplicity we measure $\varphi$ from the direction of the source. The position of any elementary scattering volume (like the small cube in Fig 1) along the line of sight, at a distance $r_0$ from the source, can be parametrized by its height $h$ above ground.

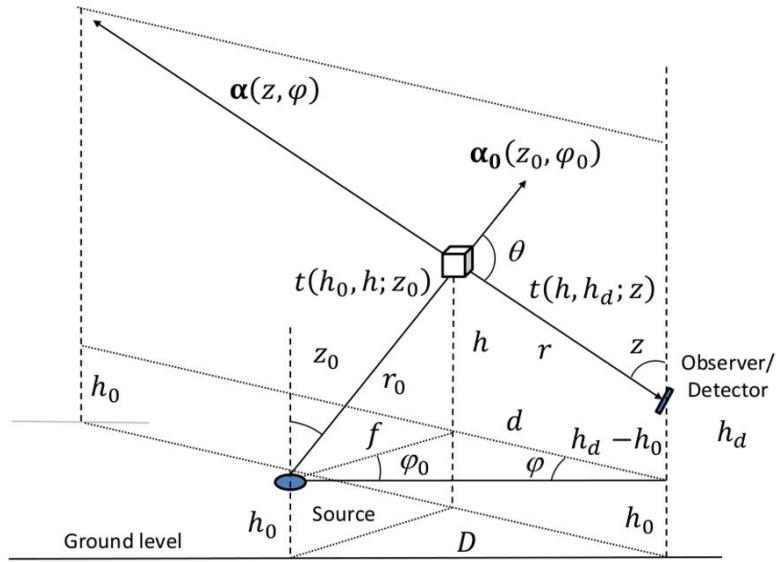

**Fig. 1.** Source and observer geometry.

No particular assumptions are made at this point about the geometrical 3-D shape or character of the source; the basic formulation below can be equally applied to a luminaire whose lamp is located few meters above ground, to a horizontal patch of road, pavement, or soil, or to a vertical patch of façade, reflecting the light from ground-level ($h_0 = 0$) or from any other arbitrary height. Given the short distances and altitudes involved in these calculations in comparison with the size of our planet, the Earth radius can be taken as infinite for all practical computations.

*2.2 Artificial sky radiance*

Let us denote by $L(\boldsymbol{\alpha_0}; \lambda)$ the spectral radiance (W·m$^{-2}$·sr$^{-1}$·nm$^{-1}$) emitted by the source in the generic direction $\boldsymbol{\alpha_0}$, defined by the angles $(z_0, \varphi_0)$, being $z_0$ the zenith angle, $\varphi_0$ the azimuth, and $\lambda$ the wavelength. In general, $z_0 \in [0, \pi]$ and $\varphi_0 \in [0, 2\pi]$, both measured in the source's reference frame, although for flat horizontal sources at ground level $z_0$ is usually restricted to the interval $[0, \pi/2]$. The source is defined 'small' in the sense that its spectral radiance $L(\boldsymbol{\alpha_0}; \lambda)$ can be considered homogeneous in all its points, and the radiance propagated from these individual points to the scattering volume will undergo substantially the same amount of atmospheric attenuation.

The artificial radiance reaching the observer from the direction $\boldsymbol{\alpha}(z, \varphi)$ is made up from the contributions of the radiances scattered by all elementary volumes located along the line of sight. The spectral irradiance $\mathrm{d}\tilde{E}_0$ (W·m$^{-2}$·nm$^{-1}$) produced by the source at the input of an elementary volume like the one depicted in Fig. 1, on a plane perpendicular to the direction of propagation $\boldsymbol{\alpha_0}$ and assuming a perfectly transparent atmosphere, would be $\mathrm{d}E_0' = L(\boldsymbol{\alpha_0}; \lambda) \, \mathrm{d}\Omega_s$, where $\mathrm{d}\Omega_s \equiv \mathrm{d}\Omega_s(r_0, \boldsymbol{\alpha_0})$ is the solid angle (in sr) subtended by the small source, viewed from the scattering volume located at a distance $r_0$ from the source. However, in a real atmosphere this irradiance is lower, due to the attenuation by absorption and scattering along the ray path from the source (at $h_0$) to the volume (at $h$), slanted by an angle $z_0$. Denoting by $t(h_0, h; z_0; \lambda)$ the atmospheric transmittance at wavelength $\lambda$ along this path, the actual irradiance at the entrance of the scattering volume, $\mathrm{d}E_0 = t(h_0, h; z_0; \lambda) \, \mathrm{d}E_0'$ will be

$$\mathrm{d}E_0 = t(h_0, h; z_0; \lambda) \, L(\boldsymbol{\alpha_0}; \lambda) \, \mathrm{d}\Omega_s(r_0, \boldsymbol{\alpha_0}) \tag{1}$$

Specifying the variables in Eq.(1) we implicitly assume an atmosphere whose properties may depend on the altitude above ground, but not on the horizontal coordinates.

Within the elementary scattering volume, the irradiance in Eq. (1) will be partially removed from the beam, due again to the complementary processes of absorption and scattering, at a rate per unit length

$$\frac{\mathrm{d}}{\mathrm{d}r_0}(\mathrm{d}E_0) = -k(h; \lambda) \, \mathrm{d}E_0 \tag{2}$$

where $k(h; \lambda)$ (in m$^{-1}$) is the atmospheric extinction coefficient at height $h$ above the ground, and $\mathrm{d}r_0$ is the differential length element along the ray. The overall extinction coefficient $k(h; \lambda)$ results from the sum of the extinction coefficients $k_i(h; \lambda)$, $i = 1, \dots, N$ of the different molecular and aerosol constituents of the atmosphere. The fraction of the irradiance attenuated by each constituent that will be scattered, not absorbed, is determined by the single-scattering albedos $\omega_i(\lambda), i = 1, \ldots, N$, that combine into an overall scattering coefficient $k_{sca}(h; \lambda) = \sum_{i=1}^{N} \omega_i(\lambda) k_i(h; \lambda)$. The total irradiance that will be scattered from the beam in the elementary atmospheric volume depicted in Fig. 1 after travelling a distance $dr_0$ along the ray will be then

$$(d^2 E_0)_{sca} = t(h_0, h; z_0; \lambda)\, L(\boldsymbol{\alpha_0}; \lambda)\, d\Omega_s(r_0, \boldsymbol{\alpha_0})\, k_{sca}(h; \lambda)\, dr_0 \qquad (3)$$

The radiant power associated with this irradiance will be scattered unequally in different directions, according to the wavelength and physical properties of the atmospheric constituents. The angular distribution of the light scattered from the elementary atmospheric volume is described by the scattering phase function $P_i(\boldsymbol{\alpha_0}, \boldsymbol{\alpha'}; \lambda)$, whose integral extended to all possible directions of scattering is equal to $4\pi$. The equations below are expressed in terms of the angular probability density function $p_i(\boldsymbol{\alpha_0}, \boldsymbol{\alpha'}; \lambda) = P_i(\boldsymbol{\alpha_0}, \boldsymbol{\alpha'}; \lambda)/4\pi$, which is essentially the density of probability that a scattered photon will propagate in the direction $\boldsymbol{\alpha'}$, per unit solid angle around $\boldsymbol{\alpha'}$, if its direction of incidence was $\boldsymbol{\alpha_0}$. It is normalized such that its integral over the $4\pi$ sr domain of all possible scattering angles is 1. The units of $p_i(\boldsymbol{\alpha_0}, \boldsymbol{\alpha'}; \lambda)$ are sr$^{-1}$, hence its multiplication by an irradiance has the dimensions of a radiance. In the situation here considered the observer is located in the direction $\boldsymbol{\alpha'} = -\boldsymbol{\alpha}$ as seen from the elementary volume (with angles $z' = \pi - z, \varphi' = \pi - \varphi$), so that the radiance $(d^2 L')_{sca}$ scattered from the volume towards the observer is:

$$(d^2 L')_{sca} = t(h_0, h; z_0; \lambda)\, L(\boldsymbol{\alpha_0}; \lambda)\, d\Omega_s(r_0, \boldsymbol{\alpha_0})\, \Gamma(\boldsymbol{\alpha_0}, \boldsymbol{\alpha}; \lambda; h)\, dr_0 \qquad (4)$$

where

$$\Gamma(\boldsymbol{\alpha_0}, \boldsymbol{\alpha}; \lambda; h) = \sum_{i=1}^{N} \omega_i(\lambda)\, k_i(h; \lambda)\, p_i(\boldsymbol{\alpha_0}, -\boldsymbol{\alpha}; \lambda) \qquad (5)$$

The radiance in Eqs. (4-5) is again attenuated along the path of length $r$ from the volume (at $h$) to the detector (at $h_d$), slanted by an angle $z$. Denoting by $t(h_d, h; z; \lambda)$ the atmospheric transmittance along this path, the spectral radiance $(d^2 L)_{sca} = t(h_d, h; z; \lambda)\, (d^2 L')_{sca}$ received at the detector from the elementary atmospheric scattering volume, due to the small source subtending a solid angle $d\Omega_s$, is

$$(d^2 L)_{sca} = t(h_0, h; z_0; \lambda)\, t(h_d, h; z; \lambda)\, L(\boldsymbol{\alpha_0}; \lambda)\, d\Omega_s(r_0, \boldsymbol{\alpha_0})\, \Gamma(\boldsymbol{\alpha_0}, \boldsymbol{\alpha}; \lambda; h)\, dr_0 \qquad (6)$$

For convenience of notation, we can make explicit the variables and parameters on which $(d^2L)_{sca}$ depends:

$$(d^2L)_{sca} \equiv d^2L_{sca}(D, \lambda, \boldsymbol{\alpha_0}, \boldsymbol{\alpha}, \boldsymbol{\beta}, h, d\Omega_s) = G_{sca}(D, \lambda, \boldsymbol{\alpha_0}, \boldsymbol{\alpha}, \boldsymbol{\beta}, h) \, d\Omega_s \, dr_0 \qquad (7)$$

where $\boldsymbol{\alpha} = (z, \varphi)$ is the direction of observation, $\boldsymbol{\beta} = \{\boldsymbol{\beta_0}, \boldsymbol{\beta_d}, \boldsymbol{\beta_{atm}}\}$ is a set of several parameter vectors characterizing the source, $\boldsymbol{\beta_0} = [h_0, L(\boldsymbol{\alpha_0}; \lambda), \ldots]$, the observer, $\boldsymbol{\beta_d} = [h_d, \ldots]$, and the state of the atmosphere $\boldsymbol{\beta_{atm}} = [k_i, \omega_i, p_i, \ldots]$, respectively, and where we have defined

$$G_{sca}(D, \lambda, \boldsymbol{\alpha_0}, \boldsymbol{\alpha}, \boldsymbol{\beta}, h) = t(h_0, h; z_0; \lambda) \, t(h_d, h; z; \lambda) \, L(\boldsymbol{\alpha_0}; \lambda) \, \Gamma(\boldsymbol{\alpha_0}, \boldsymbol{\alpha}; \lambda; h) \qquad (8)$$

which is the spectral radiance received at the detector from the elementary scattering volume, per unit solid angle of the source, $d\Omega_s$, and unit propagation length, $dr_0$. We keep the explicit dependence on $\boldsymbol{\alpha_0}$ for an easier interpretation of the formulae, although $\boldsymbol{\alpha_0}$ is determined for every value of $h$ by the remaining geometrical parameters of the source-observer configuration.

Using this short-hand notation, the total radiance at the detector is obtained after the following four integrations:

(i) The spectral radiance from the elementary volume in Fig.1 is given by the integral of Eq. (7) over the whole solid angle subtended by the source, $\Omega_s(h, \boldsymbol{\alpha_0}, \boldsymbol{\beta})$,

(ii) The spectral radiance arriving to the detector from the direction $\boldsymbol{\alpha}$ will be the integral of (i) along the whole line of sight. The integration variable can be changed from $r_0$ to $h$ recalling that $\cos z_0 = (h - h_0)/r_0$ so that $dr_0 = dh/\cos z_0$. The integral is carried out from $h = h_d$ to the limit of the atmosphere. This limit can be formally taken as $\infty$ because the effective integration domain is limited by the concentration of atmospheric scatterers via $k_{sca}(h; \lambda)$. Note that if the observer is at an altitude smaller than that of the source ($h_d < h_0$) there are atmospheric volume elements for which $\cos z_0 < 0$, since $z_0$ can be larger than $\pi/2$. In this case $h$ decreases along the direction of propagation of the ray ($dh < 0$ for $dr_0 > 0$). In order to keep $dh$ formally positive along the whole integration domain $[h_d, \infty)$, $\cos z_0$ may be replaced by its modulus, $|\cos z_0|$, in the pertinent equations.

Calculating the field-of-view averaged, in-band radiance $L(\boldsymbol{\alpha_d}; D, \boldsymbol{\beta})$ reported by a calibrated radiometer with its optical axis pointing to the direction $\boldsymbol{\alpha_d}$ of the sky requires two additional integrations, namely:

(iii) one on the sky directions, **α**, weighted by the instrument's field-of-view function $F(\boldsymbol{\alpha}, \boldsymbol{\alpha}_d; \lambda)$, which accounts for the different sentitivity of the instument to the radiance arriving from different directions around its optical axis. This function is normalized such that $\int_{\Omega_{FF}} F(\boldsymbol{\alpha}, \boldsymbol{\alpha}_d; \lambda) \, d^2\boldsymbol{\alpha} = 1$ where $d^2\boldsymbol{\alpha} = \sin z \, dz \, d\varphi$ is the solid angle element in spherical coordinates around the direction $\boldsymbol{\alpha}(z, \varphi)$. In many practical cases this function can be considered independent from $\lambda$ and angularly shift-invariant, so that $F(\boldsymbol{\alpha}, \boldsymbol{\alpha}_d; \lambda) = F(\boldsymbol{\alpha} - \boldsymbol{\alpha}_d)$. Additionally, if it is rotationally symmetric around the optical axis then $F(\boldsymbol{\alpha} - \boldsymbol{\alpha}_d) = F(\|\boldsymbol{\alpha} - \boldsymbol{\alpha}_d\|)$, where "$\| \quad \|$" stands for modulus of the vector. The integration is formally extended to $\Omega_F = 4\pi$ sr, the whole set of directions in the space, albeit it can usually be restricted to the front-facing hemisphere ($2\pi$ sr) with the polar axis oriented in the direction $\boldsymbol{\alpha}_d$.

(iv) and the remaining integration is on wavelengths, $\lambda$, across the spectral range of the source, weighted by the detector spectral sensitivity passband $S(\lambda)$.

The final result for the detected radiance is then:

$$L(\boldsymbol{\alpha}_d; D, \boldsymbol{\beta}) =$$

$$= \int_\lambda S(\lambda) \int_{2\pi} F(\boldsymbol{\alpha}, \boldsymbol{\alpha}_d; \lambda) \int_{h=h_d}^{\infty} \left[ \int_{\Omega_s} G_{sca}(D, \lambda, \boldsymbol{\alpha_0}, \boldsymbol{\alpha}, \boldsymbol{\beta}, h) \, d\Omega_s(r_0, \boldsymbol{\alpha_0}) \right] \frac{dh}{|\cos z_0|} \, d^2\boldsymbol{\alpha} \, d\lambda \quad (9)$$

The calculation of the radiance given by Eq. (9) requires the use of numerical methods, excepting for very simple, conceptually interesting canonical cases like the one described in subsection 2.4 that can be analytically described. Some useful geometrical relationships and particular expressions for an exponential atmosphere are summarized in the Appendix, sections A1 and A2, respectively.

*2.3 The solid angle of the source*

A short comment is in place regarding the integration over the solid angle subtended by the source. For small-size horizontal planar sources of area $dA_s$ observed from distances $r_0$ such that $r_0^2 \gg dA_s$ the approximation $d\Omega_s(r_0, \boldsymbol{\alpha_0}) = dA_s |\cos z_0|/r_0^2$ is valid (the use of the absolute value of the cosine allows applying this expression also to cases where $z_0 > \pi/2$).

Note however that the solid angle actually subtended by a given patch of area $dA_s$ as the distance $r_0$ tends to zero does not increase indefinitely, but tends to saturate at $2\pi$ sr. Some care shall be exercised when applying Eq.(9) for very close distances to the sources in order to compute accurately the detected radiance, avoiding formal singularities that could arise at $r_0 = 0$ if the approximate solid angle expression is not integrated correctly, especially when using discrete sums for evaluating the solid angle integral. In this regard, recall that the solid angle $\Omega_\theta$ subtended by a cone of apex angle $2\theta$ is $\Omega_\theta = 4\pi \sin^2(\theta/2)$. For a circular flat source of area $A_s$ and diameter $l$ viewed from a distance $r_0$ at the perpendicular to its center ($z_0 = 0$), the cone half-angle $\theta$ is given by

$$\theta = \tan^{-1}\left(\frac{l}{2r_0}\right) = \tan^{-1}\left(\frac{1}{r_0}\sqrt{\frac{A_s}{\pi}}\right) \tag{10}$$

When observed under a generic zenith/nadir angle $z_0$, the source cross section is reduced to $A_s|\cos z_0|$ and the solid angle becomes

$$\Omega_s(r_0, \boldsymbol{\alpha_0}) = 4\pi \sin^2\left[\frac{1}{2}\tan^{-1}\left(\frac{1}{r_0}\sqrt{\frac{A_s|\cos z_0|}{\pi}}\right)\right] = 4\pi \sin^2\left[\frac{1}{2}\tan^{-1}\left(\frac{\cos z_0}{h - h_0}\sqrt{\frac{A_s|\cos z_0|}{\pi}}\right)\right] \tag{11}$$

where we have substituted $r_0 = (h - h_0)/\cos z_0$. For large values of $r_0$ (in the sense $r_0^2 \gg A_s$) we recover from Eq. (11) the usual approximation $\widehat{\Omega}_s(r_0, \boldsymbol{\alpha_0}) = A_s|\cos z_0|/r_0^2$. When $r_0 \to 0$ Eq.(11) leads to $\Omega_s(r_0, \boldsymbol{\alpha_0}) \to 2\pi$, for any nonzero constant value of $A_s|\cos z_0|$. The approximate $\widehat{\Omega}_s$ and exact $\Omega_s$ values of the solid angle subtended by a circular patch of area $A_s$ viewed from different distances $r_0$ (with $|\cos z_0| = 1$) perpendicular to its center are displayed in Fig. 2.

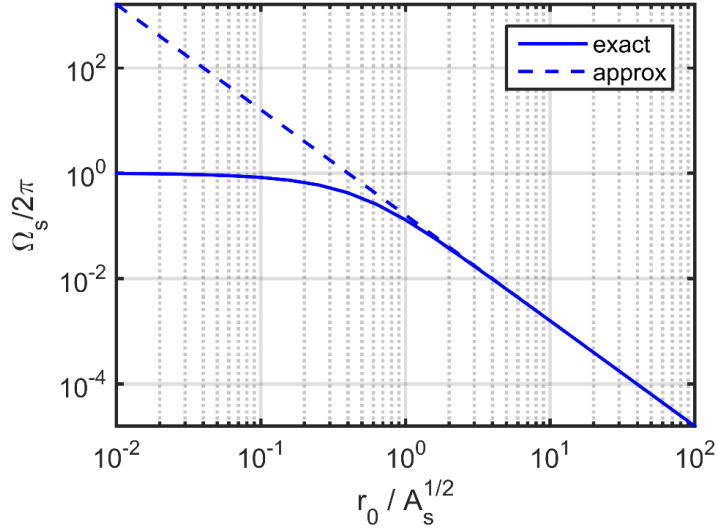

**Fig. 2.** Solid angle (in units $2\pi$ sr) subtended by a circular region of area $A_s$ viewed from a point at different distances $r_0$ perpendicular to its center, measured in units $A_s^{1/2}$. The exact ($\Omega_s$) and approximate ($\widehat{\Omega}_s$) expressions are sensibly coincident for distances $r_0 > A_s^{1/2}$. As $r_0 \to 0$ the exact expression quickly tends to $2\pi$ sr whereas the approximate one diverges. See text for details.

*2.4 Dependence of the radiance on the distance and height above ground in a simplified case*

The analysis of a simplified situation is often useful for getting some physical insights about the phenomenon under study. The basic knowledge thus achieved may be complemented later with detailed and accurate evaluations of more complex cases. Before providing in Section 3 the numerical evaluations of Eq. (9) for several configurations of interest, let us consider here one of these very simplified situations, with particular focus on the behavior of the artificial zenith sky radiance detected by an observer at short horizontal distances $D$: a small, flat, horizontal ground-level source with monochromatic Lambertian radiance within a homogeneous, finite thickness and isotropically scattering atmosphere.

As shown in detail in Appendix (A3), the artificial zenith radiance measured from an observer located at the very center of the source ($D = 0, h_d = h_0 = 0$) saturates at a finite level given by Eq. (A3.7). As the observer moves away from the source, the artificial zenith radiance decreases first slowly, until reaching the distance $D \sim \sqrt{A_s}$, and then as $D^{-1}$ within a range which typically includes the first hundreds of meters. For this range of distances, when the source and the observer are both at ground level (and under the remaining conditions stated in Appendix A3), we find from Eq.(A3.13) that

$$L(\mathbf{0};D) \approx \frac{\pi\, C_0\, A_s}{2\, D} \qquad (12)$$

where $C_0$ is a parameter independent from $D$, defined in Eq. (A3.6). Hence the decimal logarithm of the radiance decreases linearly with $\log_{10} D$, with slope -1, which means that the artificial zenith sky darkness increases by 2.5 mag/arcsec$^2$ for every log10 unit of distancing from the source. This expected reduction in the artificial sky brightness provides some rationale to the observed increase in the quality of the night sky when displacing oneself one log10 unit of $D$, for instance from 10 to 100 m away from an isolated streetlight.

## 3. Results

### *3.1. Parameters of the calculation*

A basic street lighting configuration is the one composed of a luminaire and several surrounding surfaces, in most cases horizontal (pavements) or vertical (façades). These surfaces act as secondary sources by re-emitting the light after reflections, and/or as obstacles blocking the propagation of light. The scattered artificial radiance detected at short distances from the source by an eye or a photometric instrument can be described by first-order scattering models. Within this approximation, the detected radiance is made up from three main components: (i) the direct light from the lamp that was scattered by the atmosphere; (ii) the light reflected from the surfaces that was scattered by the atmosphere; and (iii) the direct light from the lamp scattered within the eye (intraocular scattering) or the structures of the measuring instrument (stray light).

In this section we analyze the contributions (i) and (ii), leaving for future work the detailed analysis of the intraocular scattering and instrument stray light. The goal of this section is to provide quantitative insights on the behavior of the detected zenith radiance, especially on its dependence on geometrical parameters like the source and observer altitudes, $h_0$ and $h_d$, and the horizontal distance between them, $D$. Since the detected radiance due to a spatially extended source (streetlamp or surface area) can be expressed as the sum of the detected radiances produced by all small patches in which the extended source can be divided, we will focus for the purposes of this section on a small flat radiant source of area $A_s = 0.01$ m² (e.g., a 10x10 cm² square) located at $h_0 = 0$, which would correspond to a patch of pavement or to an ornamental ground recessed lamp, or at some altitude over ground, $h_0 > 0$, which would describe a flat LED luminaire installed parallel to the pavement. Since we are interested in short distance effects, we will consider observers located at horizontal distances $D = 0$ m to $D = 100$ m from the source, and at heights above ground $h_d = 0$ m to $h_d = 100$ m. The source is assumed Lambertian, with radiance independent from the angle of emission, within the range of non-blocked emission directions.

The results in this section correspond to a stratified atmosphere composed of molecules and aerosols. The molecular exponential atmosphere has a characteristic scale height $H_m = 8$ km, and its optical depth is assumed to follow a Teillet [31] wavelength dependence $\tau_0^{(M)}(\lambda) = 0.00879 \times \lambda^{-4.09}$, with $\lambda$ expressed in $\mu m$, with single scattering albedo $\omega_{(M)}(\lambda) = 1$ for all visible wavelengths. The molecular scattering angular probability density function $p_{(M)}(\theta)$ follows a standard Rayleigh law (Appendix 2, Eq. (A2.4)). The aerosol component is assumed to have a characteristic scale height $H_A = 1/0.65 \; km \approx 1.54 \; km$, with aerosol optical depth $\tau_0^{(A)}(\lambda) = \tau_0^{(A)}(\lambda_0)(\lambda/\lambda_0)^{-a}$ with $\tau_0^{(A)}(\lambda_0 = 550 \; nm) = 0.2$, Ångström exponent $a = 1$ and albedo $\omega_{(A)}(\lambda_0 = 550 \; nm) = 0.85$. The aerosol scattering angular probability density function is represented by a Henyey-Greenstein model, Eq. (A2.5), with asymmetry parameter $g(\lambda_0 = 550 \; nm) = 0.80$. The results in the subsections below correspond to a quasi-monochromatic Lambertian source emitting a nominal spectral radiance of 1.00 W·m⁻²·sr⁻¹·nm⁻¹ at $\lambda = 550 \; nm$, with bandwidth 1 nm.

*3.2. Horizontal source at ground level*

As traditional gas-discharge luminaries are progressively being replaced by LEDs, the weight of the reflections in urban surfaces as a source of light pollution increases, relative to the lamp radiance directly sent towards the sky. Most present-day LED luminaires send little to no light in directions above the horizontal (excepting when purposely oriented that way), given the essentially planar design of their printed circuit boards. Analyzing the contribution of the illuminated horizontal urban surfaces is then a useful starting point to get some insights about the radiance field in the vicinity of the lamps, when observing in the zenith direction and the observer is located outside the direct lamp illumination cone.

Figure 3(a) displays in matrix form the zenith radiance given by Eq. (9) for $h_0 = 0$ at different distances $D$ from the source (horizontal axis) and observers located at different heights $h_d$ above ground level (vertical axis). The values displayed correspond to $\log_{10}[L(D, h_d)/L_{ref}]$, the decimal logarithm of the zenith radiance for every $(D, h_d)$ combination relative to the zenith radiance that would measure an observer located precisely at the source center, $L_{ref} = L(D=0, h_d=0)$. For the situation with the parameters described above, we have $L_{ref} = 7.27 \times 10^{-7}$ W·m$^{-2}$·sr$^{-1}$·nm$^{-1}$ at $\lambda = 550\ nm$.

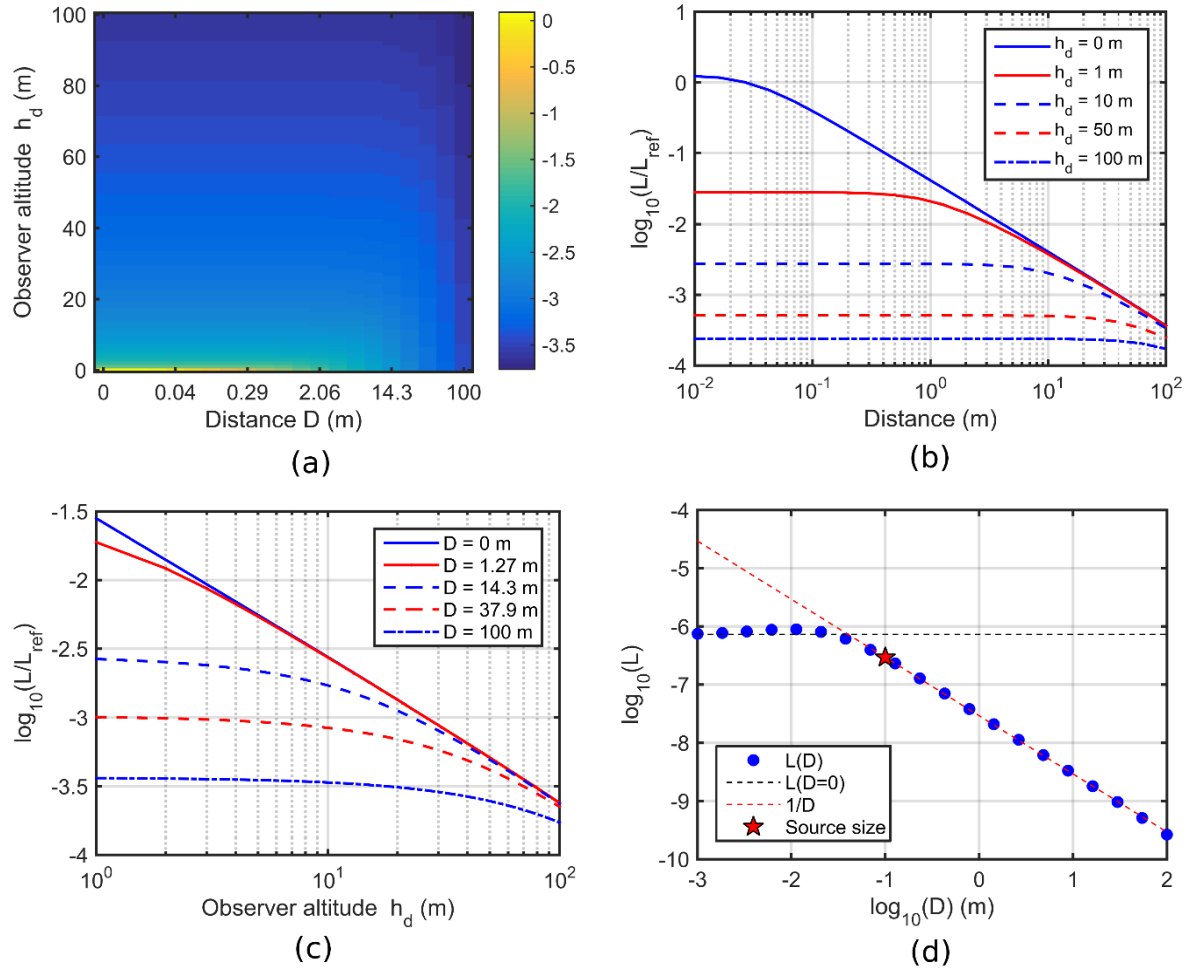

**Fig. 3.** (a): $\log_{10}[L(D, h_d)/L_{ref}]$ for different distances $D$ from the source (horizontal axis) and observers located at different heights $h_d$ above ground level (vertical axis). $L_{ref} = L(D=0, h_d=0)$ is the zenith radiance that would measure an observer located precisely

onto the source center. (b) $\log_{10}[L(D, h_d)/L_{ref}]$ as a function of $D$ for several $h_d$. (c) $\log_{10}[L(D, h_d)/L_{ref}]$ as a function of $h_d$ for several $D$. (d) The radiance computed according to the exact expression in Eq.(9) as a function of $D$ for $h_d = 0$, with superimposed approximate analytical behaviors $L \sim L(0)$ and $L \sim 1/D$, expected for $D < \sqrt{A_s}$ and $D > \sqrt{A_s}$, respectively, according to the expressions in Appendix 3 and section 2.4. The star symbol is located over the curve at $D = \sqrt{A_s}$.

The plots in Fig 3(b-c) display the dependence of the zenith radiance on $D$ (for several $h_d$) and on $h_d$ (for several $D$), respectively. As expected the detected radiance decreases for larger distances from the source (larger attenuation) and larger observer's altitude (less amount of scatterers contributing to the zenith radiance). The circles in Fig 3(d) show the values of the radiance for an observer at ground-level, $\log_{10} L(D, h_d = 0)$, which correspond to the curve $h_d = 0$ in Fig. 3(b). We have superimposed in this panel a horizontal line with the value corresponding $D = 0$, and the straight line $-\log_{10}(D)$ predicted by the simplified analytical model in Appendix 3, Eq. (A3.14). A star symbol located over the curve at the distance $D \sim \sqrt{A_s}$ illustrates the change of regime from constant radiance to radiance depending on $1/D$ as described in the Appendix and commented in section 2.4 above. Expressed in astronomical magnitudes, if any other sources of light (including natural ones) are absent or their contributions are irrelevant, this amounts to a darkening of 2.5 mag/arcsec$^2$ for every log10 unit of distancing from the source.

Figure 4 shows the contributions to the measured zenith radiance of the elementary atmospheric volumes located in the first 100 meters above ground for an observer at $h_d = 0$, at different distances from the source. These values correspond to the function $G_{sca}(D, \lambda, \pmb{\alpha_0}, \pmb{\alpha}, \pmb{\beta}, h) \, d\Omega_s(r_0, \pmb{\alpha_0})$ in Eq. (9) and have dimensions of spectral radiance produced per unit length in the vertical air column. The cumulative radiance produced by increasingly large spans of the air column, up to 10 km altitude, is shown in Fig 4(b). The integral in $dh$ in Eq. (9) was carried out up to 100 km height, but the relative contributions of the atmospheric layers above 10 km were negligible at these short distances from the source. As a matter of fact, most of the artificial zenith radiance at short distances comes from very modest altitudes, roughly equal to the distance from the source, $D$. This is quantitatively shown in Fig.(5), where the indicator $h_{50}$ (the altitude above ground of the air column responsible for the 50% of the zenith artificial radiance) is plotted versus $D$. Excepting at extremely short distances from the source center, the $h_{50}$ during the first hundred meters away from the source is practically equal, although not exactly identical, to $D$.

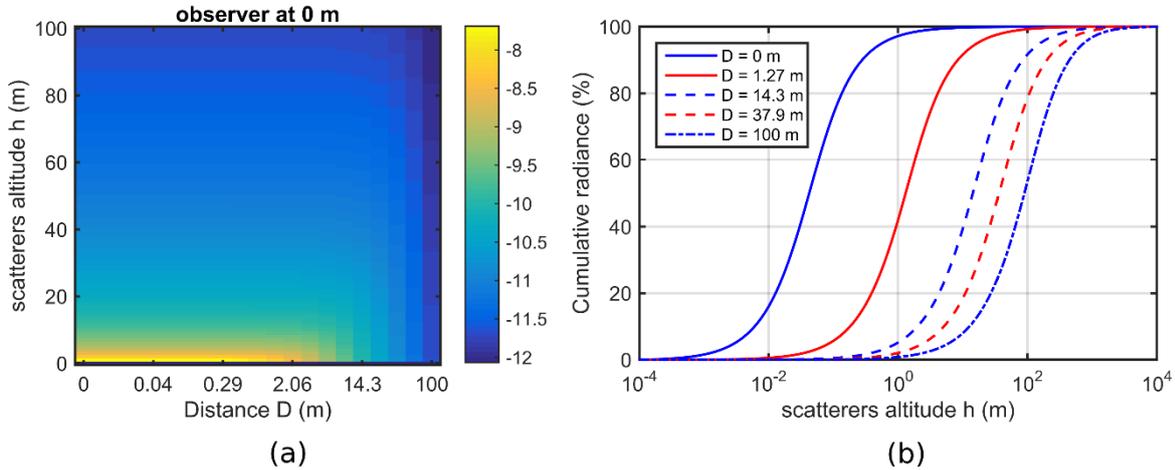

(a)  (b)

**Fig. 4.** (a): Contributions to the measured zenith radiance of the elementary atmospheric volumes located at altitudes $h$ in the first 100 meters above ground for an observer at $h_d = 0$, at different horizontal distances $D$ from the source. (b) Cumulative detected radiance produced by the vertical air column from ground to $h$, for observers at several distances $D$.

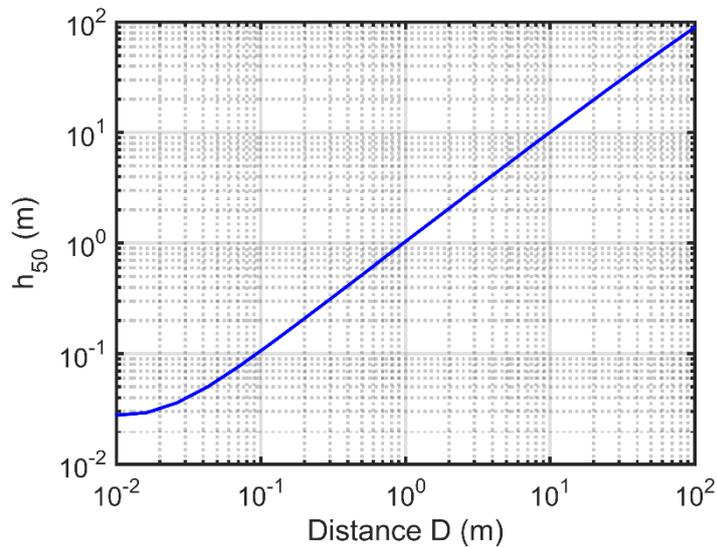

**Fig. 5.** Altitude $h_{50}$ measured from ground-level of the air column responsible for 50% of the detected zenith radiance, in the conditions analyzed in this section. During the first one hundred meters, excepting for observers very close to the center of the source $D < \sqrt{A_s}$, this altitude is practically $h_{50} \approx D$. See text for details.

*3.3. Horizontal source above ground*

We consider here a source with the same characteristics as in subsection 3.2, excepting that it is located at $h_0 = 6\ m$ above ground, and its emissions are restricted to the range $105° \leq z_0 \leq 180°$, that is, a luminaire emitting a cone of constant radiance within the range of nadir angles $[0°, 75°]$. The air column directly illuminated by this source has a decreasing limiting altitude, given by $h_{max}(D) = h_0 + D/\tan[min(z_0)]$, which in our case leads to $h_{max} = 0$ for $D = D_{max} = 22.4\ m$. This means that ground-level observers at distances longer than $D_{max}$ will detect no zenith light from the first-scattering of source photons (although, of course, they will still detect the scattered photons from light reflected on the ground and other surfaces, higher-order atmospheric scattering and natural sources in the sky). The same will happen to observers at $D < D_{max}$ if they are located at heights $h_d > h_{max}(D)$. Recall that we are analyzing here the atmospheric scattered light, not the direct zenith radiance from the source when the observer is located just below it ($D < \sqrt{A_s}$), which is of course much larger than the scattered one.

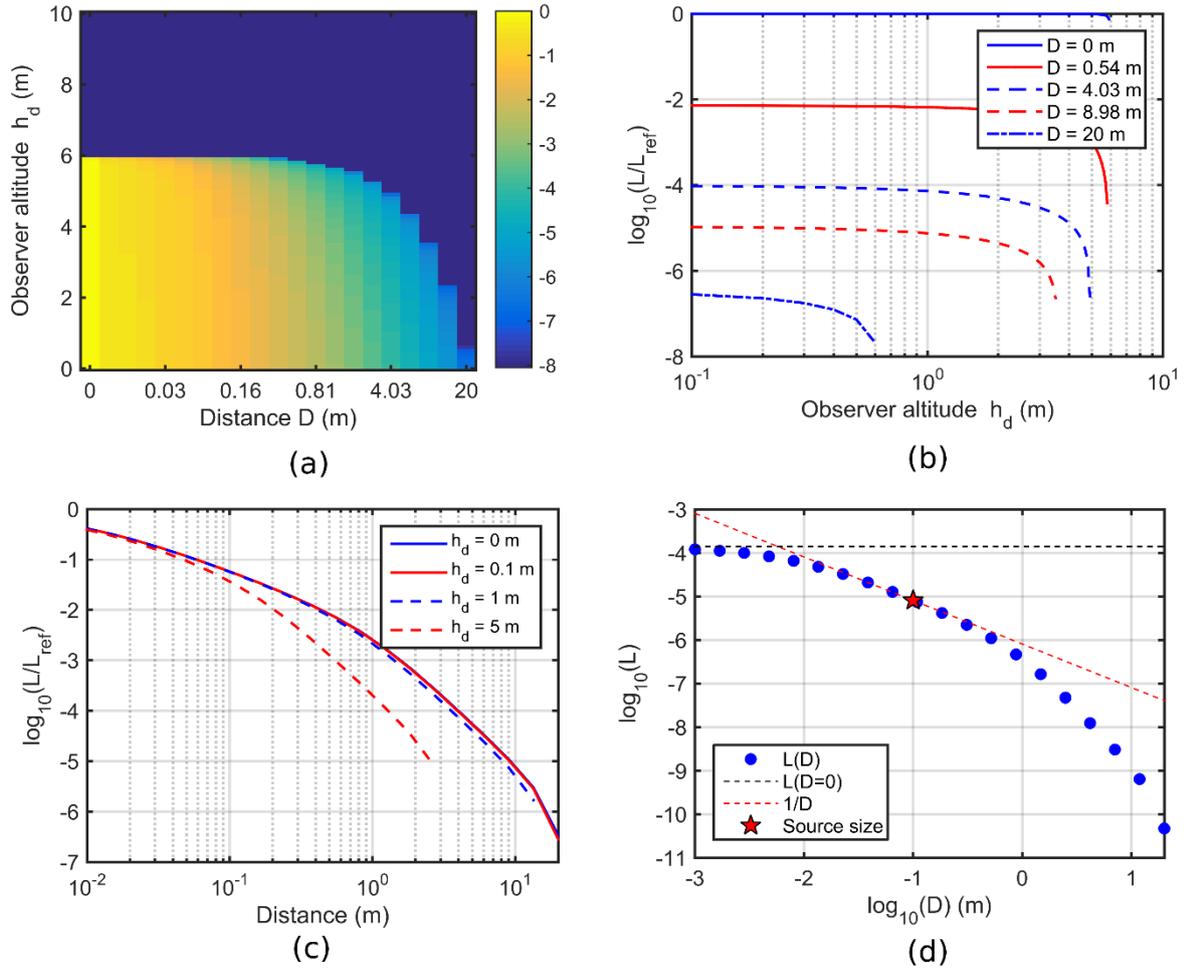

**Fig. 6.** Plots for $\log_{10}[L(D,h_d)/L_{ref}]$ as in Fig.3(a)-(d), but for a horizontal flat Lambertian source located at $h_0 = 6\,m$ above ground, emitting in angles restricted to the range $105° \leq z_0 \leq 180°$, that is nadir angles $\in [0°, 75°]$. $L_{ref} = L(D=0, h_d=0)$ is the scattered zenith radiance that would measure an observer located at ground level beneath the source center (only the scattered radiance is included in $L(D,h_d)$ and $L_{ref}$, not the direct radiance from the source). The range of displayed distances is restricted to $D = 20\,m$, since for $D > D_{max} = 22.4\,m$ the luminaire does not directly illuminate any segment of the air column.

Figure 6 displays the results for this case, with the same meaning as in Fig.3. The most relevant differences come from the fact that the illuminated volume of the atmosphere is now strongly restricted by the luminaire emission angles, oriented toward ground. This leads to no first-order scattered radiance produced above the non-illuminated atmospheric volume, and consequently a different dependence of the detected radiance with $(D, h_d)$ in comparison with Fig.3. And, as shown in Fig 6(d), the radiance $L(D, 0)$ for a ground-level observer shows a less marked correspondence with the constant and $1/D$ dependences of the radiance deduced for a ground-level source with Lambertian emission towards the whole upper hemisphere (Fig 3(d)). These differences are also noticeable in the distribution of scattered radiance throughout the atmospheric volumes and the associated $h_{50}$ indicator, Fig. (7).

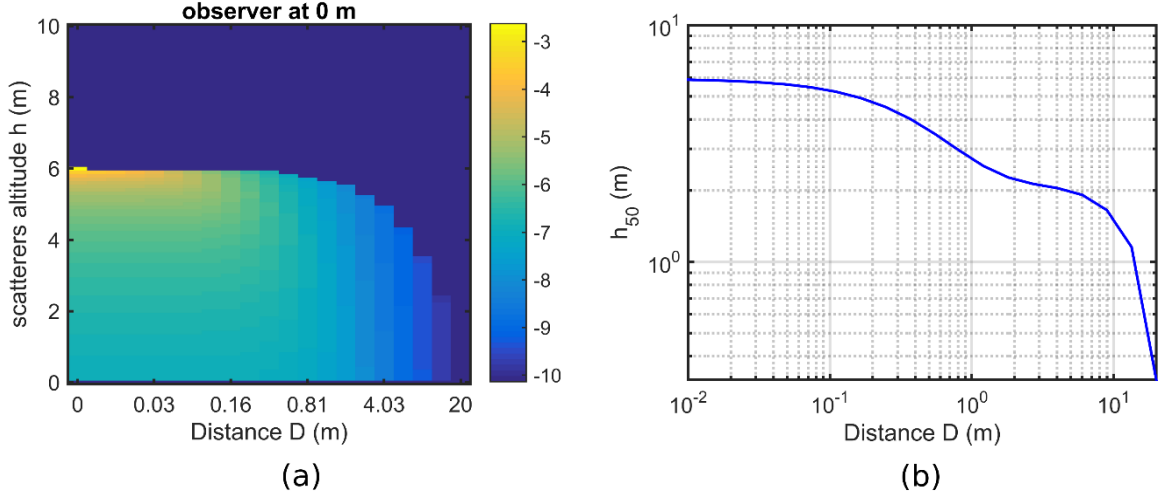

**Fig. 7.** (a): Contributions to the measured zenith radiance of the elementary atmospheric volumes located at altitudes $h$ in the first 10 meters above ground for an observer at $h_d = 0$, at different horizontal distances $D$ from the same source as in Fig. 6, located at $h_0 = 6\,m$. (b) Altitude $h_{50}$ measured from ground-level of the air column responsible for 50% of the detected zenith radiance.

*3.4. The effect of obstacles*

Let us finally consider the effect of obstacles. We choose a flat horizontal source as in subsection 3.3, located at $h_0 = 6\,m$ but with full Lambertian emission profile for all angles in its upper and lower hemisphere, such that its radiance is constant in the range $0° \leq z_0 \leq 180°$. Let us additionally asume the existence of a non-reflecting obstacle of height $h_{obst} = 7\,m$, located a distance $D_{obst} = 4\,m$ away from the source. The net effect of this obstacle is increasing the actual lower limit of integration in $dh$ in Eq.(9) when the observer is located in the shadow of the source, a situation arising when $D > D_{obst}$ and $h_d < h_{lower}$, being $h_{lower} =$ $h_0 + [(h_{obst} - h_0)/D_{obst}]D$ the lower end of the air column directly illuminated by the source in the region $D > D_{obst}$.

Given the assumed null reflectance of the object, it is expected that the behavior of the zenith radiance $L(D, h_d)$ for $D < D_{obst}$ will share some similarities with the previosly analyzed cases (with differences arising from the fact that the present case is not a mere linear combination of the previous ones). For $D > D_{obst}$ and $h_d < h_{lower}$ the similarities are expected to be greater with the case analyzed in subsection 3.2 (different altitude of the sources aside). This can be observed in Figs. (8)-(9), that show two different regimes of behavior coarsely delimited by $D_{obst} = 4\ m$.

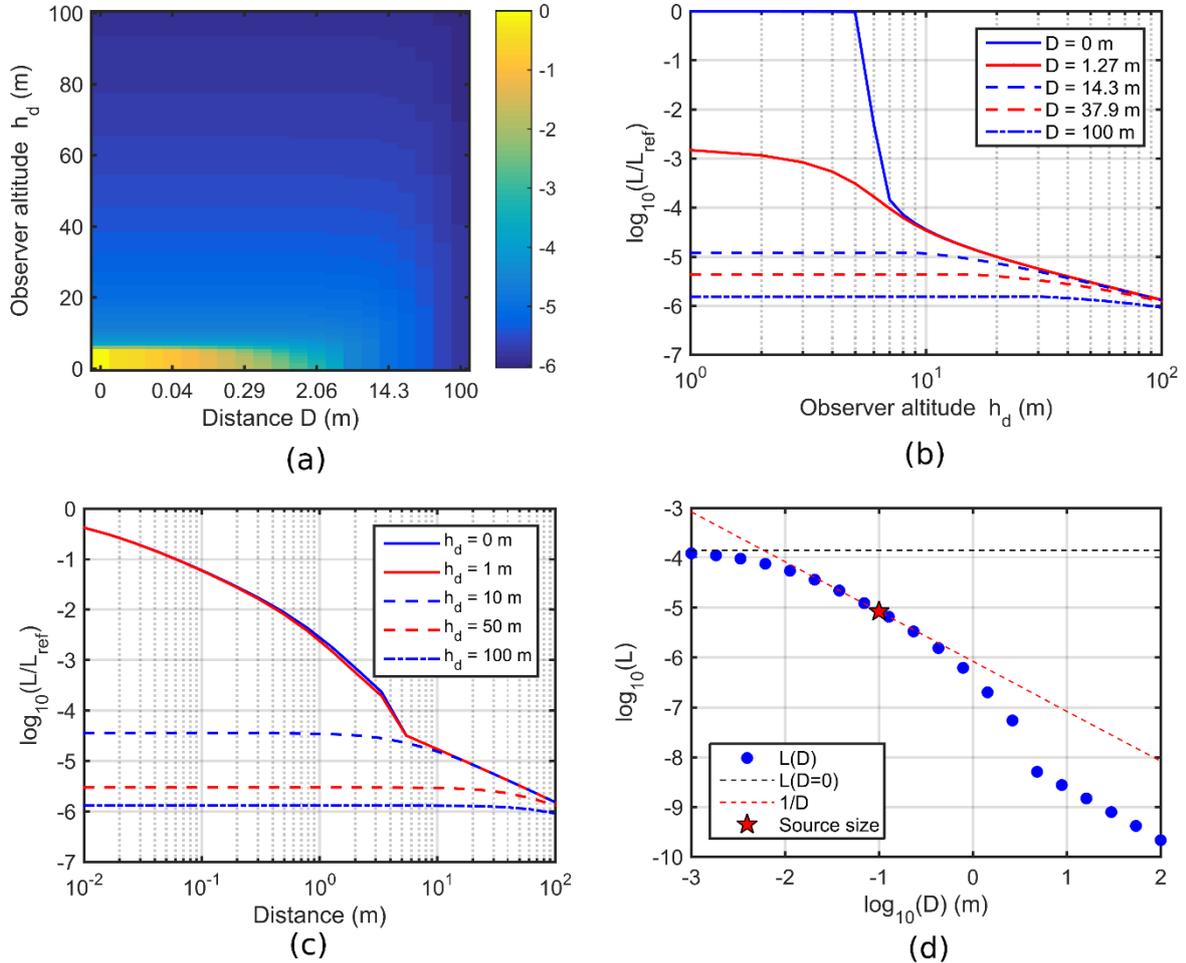

**Fig. 8.** Plots for $\log_{10}[L(D, h_d)/L_{ref}]$ as in Figs.(3,6) (a)-(d), but for a horizontal flat Lambertian source located at $h_0 = 6\ m$ above ground, emitting in the full range of angles $0° \leq z_0 \leq 180°$, with a non-reflecting obstacle of height $h_{obst} = 7\ m$, located a distance $D_{obst} = 4\ m$ away. $L_{ref} = L(D = 0, h_d = 0)$ is the scattered zenith radiance that would measure an observer

located at ground level beneath the source center (only the scattered radiance is included in $L(D, h_d)$ and $L_{ref}$, not the direct radiance from the source).

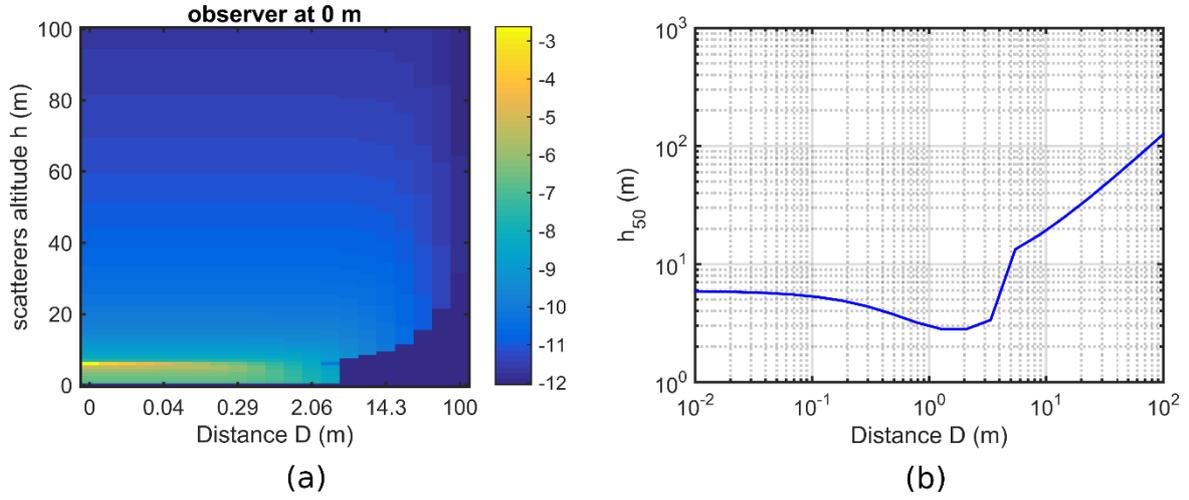

**Fig. 9.** (a): Contributions to the measured zenith radiance of the elementary atmospheric volumes located at altitudes $h$ in the first 100 meters above ground for an observer at $h_d = 0$, at different horizontal distances $D$ from the same source as in Fig. 8, located at $h_0 = 6\,m$ and with with a non-reflecting obstacle of height $h_{obst} = 7\,m$, located a distance $D_{obst} = 4\,m$ away. (b) Altitude $h_{50}$ measured from ground-level of the air column responsible for 50% of the detected zenith radiance. There is actually a sharp discontinuity of $h_{50}$ at $D = 4\,m$, the softer slope in the figure is due to the discrete sampling of distances.

*3.5. Nearby streetlights versus surrounding city lights*

Assessing the relative contribution of a nearby streetlight to the artificial zenith brightness seen by an observer, in comparison with the brightness due to the ensemble of lights of the surrounding city, requires setting the conditions of the problem. We provide in this subsection an order-of-magnitude estimate, based on realistic conditions and using some additional assumptions. To that end, let us consider an observer located in the center of a city of radius $R$ km, whose density of light emissions towards the city surfaces is $E_{in}$ in units W·m⁻² within the CIE $V(\lambda)$ photopic spectral sensitivity band [32], corresponding to a luminous density of emissions $E_{V,in} = 683\, E_{in}$ in lumen per square meter. $E_{in}$ and $E_{V,in}$ are the city-averaged irradiance and illuminances, respectively, not to be confounded with the actual average irradiance/illuminance on the streets, the latter being given by $E_{in}/\epsilon$ and $E_{V,in}/\epsilon$, where $\epsilon$ is the fraction of the city territory actually illuminated, i.e. excluding roofs and other zones where no direct light arrives from the streetlamps [28]. Assuming as a first approximation Lambertian reflections at the city surfaces, the average radiance of the city surface is $L_{city} = \rho\, E_{in}/\pi$, in units W·m$^{-2}$·sr$^{-1}$, where $\rho$ is the average ground reflectance.

Let us consider an observer, $h_d = 170$ cm tall, at a distance $D$ from a streelight whose flat emitting surface of area $A_s = 0.01$ m² is located at $h_0 = 6$ m above ground, with constant radiance within a nadir emission cone of half-angle $\gamma = 75°$, as in section 3.3. This emission cone corresponds to a solid angle $\Omega_\gamma = 4\pi \sin^2(\gamma/2)$ sr, and illuminates an area on the street $S = \pi(h_0 \tan \gamma)^2$ m². Assuming this lamp provides to the illuminated area an irradiance $E_{in}/\epsilon$, its radiant flux is $\Phi = S\, E_{in}/\epsilon$ W, and the radiance of its emitting surface is $L_{lamp} = \Phi/(\Omega_\gamma A_s)$ in W·m$^{-2}$·sr$^{-1}$.

The artificial sky brightness perceived by the observer is the sum of two terms: the radiance produced by the ensemble of all city lights scattered along the air column above the observer, from $h = h_d$ up to the limits of the atmosphere, and the radiance of the lamp scattered in the segment of the air column from the eye of the observer $h = h_d$, up to the upper limit of the lamp emission cone, $h = h_{max}(D) = h_0 - D/\tan \gamma$. The observer is outside this illuminated cone when the distance from the base of the streelight is such that $h_d > h_{max}(D)$. In this case the observer only perceives the scattered light due to the ensemble of the city surfaces (including the one illuminated by the nearby streetlight).

The radiance of the sky can be expressed in astronomical units magnitudes per square arcsecond, mag$_V$/arcsec², within the Johnson-Cousins V band [33,34] by means of the transformation:

$$L[mag_V/arcsec^2] = -2.5 \log_{10}\left(\frac{L[Wm^2 sr^{-1}]}{L_r}\right) \tag{13}$$

where $L_r$ is the reference radiance, or radiometric zero point, associated with the astronomical photometric band, which for the Johnson-Cousins V in the Vega +0.03 scale is $L_r = 143.168$ W·m$^{-2}$·sr$^{-1}$ [35,36]. Note that due to the minus sign in the definition, brighter skies correspond to lower mag$_V$/arcsec².

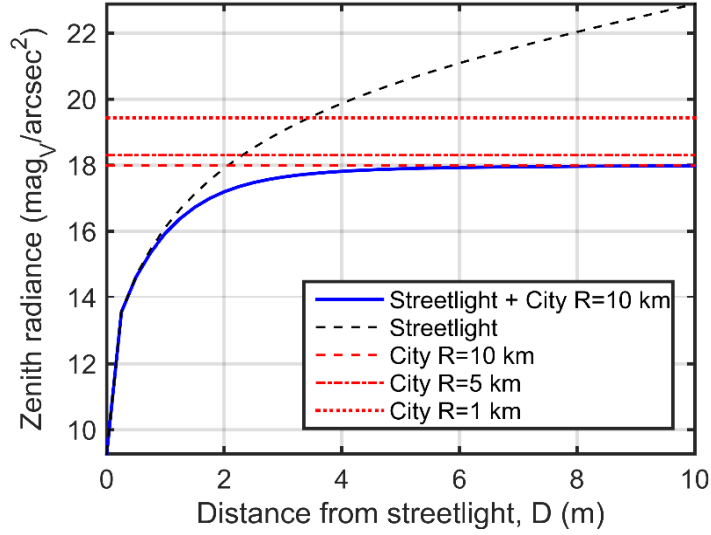

**Fig. 10.** Artificial zenith radiance mag$_V$/arcsec$^2$ perceived by an observer at horizontal distance $D$ from a streetlight surrounded by a light-emitting city, in the conditions described in the text.

For the calculations made in this example, we adopted an average city emissions density of $E_{V,in} = 2.0$ lm·m$^{-2}$ with a city factor $\epsilon = 0.5$. The calculations were made for the wavelength $\lambda = 550\ nm$, very close to the maxima of both the CIE $V(\lambda)$ and the Johnson-Cousins V. The results of the calculation are shown in Fig. 10, where the full line describes the artificial zenith radiance, in mag$_V$/arcsec$^2$, perceived by the observer at different distances from the streetlight, resulting from the combined scattered light from the streetlight itself and the whole of the city lights, assuming the observer is located at the center of a city of radius *R*=10 km. The individual contributions of the streetlight and the city are also plotted as dotted lines, the latter also including the expected values for cities of smaller size (5 and 1 km, respectively). The figure shows that up to a distance $D \approx 2$ m from the base of the lightpole the scattered radiance in the small segment of the air column above the observer (less than 4 m long) is larger than the scattered light produced by the city in the whole air column. The total radiance is still noticeably different from the one produced by the city alone until a distance $D \approx 5$ m. From this point on, the sky brightness is dominated by the baseline brightness produced by the whole city. The eyes of the observer remain within the illuminated cone of the streetlight up to a distance $D \approx 16$ m. For cities of smaller sizes with the same emission surface density the influence of the streetlamp would be noticeable at larger distances, and the resulting total brightness lines would converge towards the corresponding horizontal dotted levels shown in the Figure.

## 4. Discussion

The examples described in section 3 are meant to provide some basic insights about the expected behavior of the artificial zenith radiance in the vicinity of the urban and rural sources. Of course, they are not a comprehensive treatment of all possible situations. We have not addressed in this paper the effects of intraocular scattering and instrumental stray light, which can be highly relevant at short distances from the source when the direct source radiance enters the eye, especially at small angles with respect to the line of sight. Additionally, the luminance adaptation state of the eye determines several basic parameters that shape the visual appearance of the starry sky, among them the absolute luminance threshold and the luminance contrast threshold for detecting a star over the sky background. Given the physiological and perceptual aspects involved, this will be subject for further work.

The examples above are also restricted to zenith observation with extremely narrow field of view and quasi-monochromatic radiation at $\lambda = 550\ nm$. Given the linear character of Eq. (9), the radiance for wider fields of view and spectral power distributions can be easily calculated by performing the corresponding integrals in $\mathrm{d}^2\boldsymbol{\alpha}$ and $\mathrm{d}\lambda$. The same applies to observations made in sky directions different from the zenith one. Furthermore, for the sake of brevity we have not included here examples of the effects produced by planar vertical active sources (LED displays, windows, etc) or reflections in vertical surfaces (façades, vegetation walls, etc). The treatment of these emitters is not substantially different from the horizontal ones, excepting for the fact that the solid angle subtended by them, Eq.(11), should include now a factor $|\sin z_0|$ instead of $|\cos z_0|$.

Equation (9) provides the basic expression to compute the first-order scattered radiance produced for all these configurations by a small source. When multiple sources are at play, surrounded by extended reflecting surfaces, the resulting observed radiance can be computed by adding the individual emitter patch contributions, resorting to the whole conceptual and calculation toolbox provided by linear systems theory, as described in detail elsewhere [2,27-29,37]. The same can be said of the computation of integral indicators like, e.g., the average sky radiance or the horizontal irradiance [30,38]. The use of a single-scattering approach is justified for our problem under most practically interesting atmospheric conditions, since the short source-detector distances here considered translate into a low probability of double scattering events. The errors derived from ignoring second and higher scattering orders in artificial skyglow calculations have been comprehensively studied by Kocifaj [5]. As a typical result, the contribution of the second-scattering order for an atmosphere with aerosol optical depth or order 0.2 illuminated by an artificial source at ground level is more than one order of magnitude smaller than the contribution of the first order, for distances shorter than 1 km.

The results obtained in this work strongly suggest that there is indeed a rationale supporting the observations that the artificial sky brightness may vary noticeably in urban settings (loosely evaluated by the number of visible stars) if the observing place is displaced by a few tens or hundred meters and the relative geometry of the sources, surfaces and obstacles does change significantly. This result opens the way for designing urban microspaces where the deleterious effects of light pollution on the observation of the night sky be partially attenuated. Of course, local modifications made in special places do not reduce the baseline sky brightness caused by other multiple artificial sources installed in the surrounding territory, at distances of tens or hundreds of km away from the observer, a problem whose management requires an effective territorial planning strategy. But achieving a reduction of order 1-2 mag/arcsec$^2$ of the artificial brightness of the sky by adopting local-level measures seems to be still a useful possibility.

## 5. Conclusions

A quantitative analytical model for the artificial sky radiance in the vicinity of outdoor light sources is developed in this paper. The model is free from singularities at the origin and is useful for designing urban dark sky spaces. The analysis is restricted to first-order scattering, an approximation acceptable for this range of distances in non extremely thick atmospheres. We found that the artificial zenith sky brightness produced by a small, ground level, horizontal source at short distances (from the typical dimension of the source up to several hundred meters) depends on the inverse of the distance to the source if observer and source are located at about the same altitude. This is equivalent to a reduction of the artificial sky brightness of 2.5 mag/arcsec² for every $\log_{10}$ unit increase of the distance. The behavior of the detected scattered radiance with the altitude of the source, angular emission pattern, altitude of the observer, and screening by obstacles are also analyzed.

**Appendix**

**A.1 Basic geometrical relations**

For evaluating the integrals in Eq.(9) of section 2.1, and in particular the integral along the line of sight it is convenient to express the different quantities in terms of the geometrical parameters contained in the vectors $\boldsymbol{\alpha}, \boldsymbol{\beta}$, of $D$ and of the height above ground, $h$. According to the notation in Figure 1:

$$\cos z_0 = \frac{h - h_0}{\sqrt{D^2 + [(h - h_d) \tan z]^2 - 2D(h - h_d) \tan z \cos \varphi + (h - h_0)^2}}$$

$$\cos z = \frac{h - h_d}{r}$$

The scattering angle $\theta$ (angle between the unit vectors $\boldsymbol{\alpha_0}, -\boldsymbol{\alpha}$) is

$$\cos \theta = \frac{D^2 + (h_d - h_0)^2 - r_0^2 - r^2}{2 r_0 r}$$

where $r$ and $r_0$ are given by the basic relationships

$$\cos z = \frac{h - h_d}{r} \quad \rightarrow \quad r = \frac{h - h_d}{\cos z}$$

$$\cos z_0 = \frac{h - h_0}{r_0} \quad \rightarrow \quad r_0 = \frac{h - h_0}{\cos z_0}$$

The scattering angle is generally undetermined for $r_0 = 0$. For $r = 0$, i.e. for the lower limit of the integral along the line of sight, $h = h_d$, is value can be calculated as

$$\cos \theta \, [r = 0] = -[\cos z_0 \cos z + \sin z_0 \sin z \cos(\pi - \varphi)]$$

For zenith observations, in the general case ($\cos z = 1, r = h - h_d, r_0 = \sqrt{D^2 + (h - h_0)^2}$) we have $\cos \theta = -\cos z_0$, hence $\theta = \pi - z_0$.

**A.2 Particular expressions for the atmospheric constituents**

The numerical calculations in section 3 are carried out for a simplified layered atmosphere, composed of molecular and aerosol constituents with exponential concentration profiles. Each atmospheric constituent is characterized by its extinction coefficient $k_i(h;\lambda)$, $i = 1, \ldots, N$, associated atmospheric optical depth $\tau_0^{(i)}(\lambda)$, single-scattering albedo $\omega_i(\lambda)$, and scattering angular probability density function $p_i(\theta;\lambda)$.

For every individual atmospheric component, the transmittance along a slanted path at an angle $\zeta$ with the zenith ($\zeta = z, z_0, \ldots$) between two atmospheric altitudes $h_1, h_2$ is

$$t_i(h_1, h_2, \zeta; \lambda) = \exp\left[-M_i(\zeta) \int_{h_1}^{h_2} k_i(h'; \lambda) \mathrm{d}h'\right] = \exp[-M_i(\zeta)\, \tau^{(i)}(h_1, h_2; \lambda)] \quad (A2.1)$$

where $M_i(\zeta)$ is the airmass number, that in a first approximation can be considered independent of $\lambda$ and given by $M(\zeta) = 1/\cos\zeta$. If the expression $(A2.1)$ is applied to the whole atmosphere ($h_1 = 0, h_2 = \infty$), a correction shall be made to this expression of $M(\zeta)$ in order to avoid a strong divergence with the actual values for a spherical Earth at angles close to the horizon ($\zeta \in [\sim 75°, 90°]$), see e.g., Kasten and Young [39]. In the equation above $\tau^{(i)}(h_1, h_2; \lambda)$ is the optical depth contributed by the *i*-th component along the vertical path between $h_1, h_2$, which is defined as

$$\tau^{(i)}(h_1, h_2; \lambda) = \int_{h_1}^{h_2} k_i(h'; \lambda) \mathrm{d}h' \quad (A2.2)$$

The atmospheric optical depth contributed by the *i*-th component is given by $\tau_0^{(i)}(\lambda) \equiv \tau^{(i)}(0, \infty; \lambda)$. Since the total extinction factor due to $N$ not interacting, independent components is $k(h; \lambda) = \sum_{i=1}^{N} k_i(h; \lambda)$ it immediately follows $\tau(h_1, h_2; \lambda) = \sum_{i=1}^{N} \tau^{(i)}(h_1, h_2; \lambda)$, and $t(h_1, h_2, \zeta; \lambda) = \prod_{i=1}^{N} t_i(h_1, h_2, \zeta; \lambda)$. If the airmass dependence on the zenith angle $\zeta$ is the same for all atmospheric constituents, $M_i(\zeta) = M(\zeta)$, then $t(h_1, h_2, \zeta; \lambda) = \exp[-M(\zeta)\, \tau(h_1, h_2; \lambda)]$.

The altitude profile of the extinction coefficient for an exponential atmosphere is $k_i(h; \lambda) = k_{0i}(\lambda) \exp(-h/H_i)$. According to Eq. (A2.2), in this case $\tau_0^{(i)}(\lambda) = H_i\, k_{0i}(\lambda)$, and

$$\tau^{(i)}(h_1, h_2; \lambda) = k_{0i}(\lambda) \int_{h_1}^{h_2} \exp(-h/H_i)\, dh' = \tau_0^{(i)}(\lambda)\left[e^{-h_1/H_i} - e^{-h_2/H_i}\right] \quad (A2.3)$$

The wavelength dependence of the atmospheric optical depth for the molecular constituents $\tau_0^{(M)}(\lambda)$, can be described using a classical Rayleigh power law like, see e.g. Teillet (1990) $\tau_0^{(M)}(\lambda) = 0.00879 \times \lambda^{-4.09}$, with $\lambda$ expressed in $\mu m$. The molecular single scattering albedo $\omega_{(M)}(\lambda)$ can be taken practically equal to 1 for all visible wavelengths. The Rayleigh scattering angular probability density function (normalized to 1 under integration to all directions of scattering) is

$$p_{(M)}(\boldsymbol{\alpha_0}, -\boldsymbol{\alpha}; \lambda) = p_{(M)}(\theta) = \frac{3}{16\pi}(1 + \cos^2\theta) \quad (A2.4)$$

The wavelength dependence of the aerosol optical depth can be described, relative to some reference wavelength $\lambda_0$, by means of the Ångström exponent $a$, such that $\tau_0^{(A)}(\lambda) = \tau_0^{(A)}(\lambda_0)(\lambda/\lambda_0)^{-a}$. For the applications in this work the rigorous aerosol scattering angular probability density function can be replaced by a normalized Henyey-Greenstein function with asymmetry parameter $g$

$$p_{(A)}(\boldsymbol{\alpha_0}, -\boldsymbol{\alpha}; \lambda) = p_{(A)}(\theta; \lambda) = \left(\frac{1}{4\pi}\right)\frac{1 - g^2}{(1 + g^2 - 2g\cos\theta)^{3/2}} \quad (A2.5)$$

The wavelength dependence of the aerosol albedo $\omega_{(M)}(\lambda)$ and asymmetry parameter $g(\lambda)$ can be described by e.g. Eqs.(6-8) of McComiskey et al [40].

**A.3 Dependence of the artificial zenith brightness on the observer's distance and height above ground in a simplified configuration**

Let us here consider the behavior of the artificial zenith brightness produced by a small flat source of area $A_s$, located horizontally at ground-level ($h_0 = 0$), detected by an observer at height $h_d$ located an horizontal distance $D$ away from the source.

Isolating the radiance arriving strictly from the zenith direction ($\boldsymbol{\alpha} = \boldsymbol{0}$) is equivalent to using a point-like field of view function $F(\boldsymbol{\alpha}, \boldsymbol{\alpha}_d; \lambda) = F(\boldsymbol{0}, \boldsymbol{0}; \lambda)\delta(\boldsymbol{\alpha})$, where $\delta(\boldsymbol{\alpha})$ is the symbol for a Dirac-delta distribution centered at $\boldsymbol{\alpha} = \boldsymbol{0}$. Under these conditions, we have

$$\cos z_0 = \frac{h}{\sqrt{D^2 + h^2}} \tag{A3.1}$$

$$\Omega_s(h, \boldsymbol{\alpha_0}) = 4\pi \sin^2\left[\frac{1}{2}\tan^{-1}\left(\frac{\cos z_0}{h}\sqrt{\frac{A_s|\cos z_0|}{\pi}}\right)\right] \tag{A3.2}$$

Restricting the study to the radiance detected within an infinitesimally narrow spectral interval around an arbitrary wavelength $\lambda_0$ is equivalent to choosing a detector passband $S(\lambda) = S(\lambda_0)\, \delta(\lambda - \lambda_0)$. Denoting $S(\lambda_0)\, F(\mathbf{0},\mathbf{0};\lambda_0) \equiv S_0\, F_0$ and performing the integrations in $\lambda$, $\boldsymbol{\alpha}$, and the source solid angle in Eq.(9) with these assumptions one gets

$$L(\mathbf{0}; D, \boldsymbol{\beta}) = S_0\, F_0 \int_{h=h_d}^{\infty} \Omega_s(h, \boldsymbol{\alpha_0})\, G_{sca}(D, \lambda, \boldsymbol{\alpha_0}, \mathbf{0}, \boldsymbol{\beta}, h)\, \frac{\sqrt{D^2 + h^2}}{h}\, dh \tag{A3.3}$$

As an additional simplification, let us consider an homogeneous atmosphere with constant concentration of molecules and aerosols from $h = 0$ up to a maximum height $h = H$. Such an atmosphere has constant extinction and scattering coefficients $k(h; \lambda_0) = k$ and $k_{sca}(h; \lambda) = k_{sca}$, respectively, for $h \leq H$ (and zero otherwise). For this atmosphere and geometrical configuration, the transmittance factors are given by:

$$t(h_0 = 0, h; z_0; \lambda) = \exp\left[-k\sqrt{D^2 + h^2}\right] \tag{A3.4}$$

$$t(h_d, h; z = 0; \lambda) = \exp[-k(h - h_d)] \tag{A3.5}$$

Finally, let us assume that the effective angular scattering is overally isotropic, that is, that the product of the angular dependences of the emitted radiance and the scattered light is approximately constant for all values of $\boldsymbol{\alpha_0}$, a condition that could be fulfilled, among others, by the combination of a Lambertian source $L(\boldsymbol{\alpha_0}; \lambda) = L_0$, and an isotropically scattering atmosphere, $\Gamma(\boldsymbol{\alpha_0}, \boldsymbol{\alpha}; \lambda_0; h) = k_{sca}/(4\pi)$. Equation (A2.3) becomes then

$$L(\mathbf{0}; D) = C_0 \int_{h=h_d}^{H} \Omega_s(h, D)\, \exp\left[-k\left(h - h_d + \sqrt{D^2 + h^2}\right)\right] \frac{\sqrt{D^2 + h^2}}{h}\, dh \tag{A3.6}$$

where $C_0 = S_0\, F_0\, L_0\, k_{sca}/(4\pi)$.

Let us analyze the behavior of the radiance (A3.6) across several ranges of $D$. For $D = 0$ (observer located above the center of the source), $\cos z_0 = 1$ for all $h$, and the zenith radiance is

$$L(\mathbf{0};0) = 4\pi\, C_0 \int_{h=h_d}^{H} \sin^2\left[\frac{1}{2}\tan^{-1}\left(\frac{1}{h}\sqrt{\frac{A_s}{\pi}}\right)\right] \exp[-k(2h - h_d)]\, dh \quad (A3.7)$$

that shall be evaluated numerically, depending on the input parameters. The integral does not diverge even if the observer is located precisely at the center of the source ($h_d = 0$), since $h$ is always larger than $h_d$, and for $h = 0$ the integrand equals $(1/2)\exp(-2kh)$.

For increasing values of $D$ the solid angle term of the integral (A3.6) behaves as shown in section 2.2 (Fig. 2), first desaturating the $2\pi$ value at the origin and then decreasing with a $|\cos z_0|/r_0^2$ dependence, according to the expression $\widehat{\Omega}_s(r_0, \boldsymbol{\alpha}_0) = A_s|\cos z_0|/r_0^2$, valid for $r_0^2 > A_s$. In the geometry here considered, $|\cos z_0|/r_0^2 = h/(D^2 + h^2)^{3/2}$ so that for this range of $D$ (A3.6) has the form

$$L(\mathbf{0}; D) = C_0\, A_s \int_{h=h_d}^{H} \frac{\exp\left[-k\left(h - h_d + \sqrt{D^2 + h^2}\right)\right]}{D^2 + h^2}\, dh \quad (A3.8)$$

Performing the change of variable $h = xD$, for every $D$, we get

$$L(\mathbf{0}; D) = C_0\, A_s \left(\frac{1}{D}\right) \int_{x=x_d}^{x_H} \frac{\exp\left[-kD\left(x - x_d + \sqrt{1 + x^2}\right)\right]}{1 + x^2}\, dx \quad (A3.9)$$

where $x_d = h_d/D$ and $x_H = H/D$. For $D$ sufficiently small (but still compatible with the condition $D^2 + h^2 > A_s$ that makes it possible the use of $\widehat{\Omega}_s$ instead of $\Omega_s$ for all $h$ in the integration interval) the exponential in the integrand can be approximated by 1. This happens if

$$kD\left(x - x_d + \sqrt{1 + x^2}\right) \ll 1 \quad (A3.10)$$

for all $x \in [x_d, x_H]$. The most demanding condition on $D$ in Eq. (A3.10) arises for $x \in x_H$, which is equivalent to

$$D \ll \sqrt{\left[\frac{1}{k} - (H - h_d)\right]^2 - H^2} \quad (A3.11)$$

which, for typical values $H = 8$ km, $h_d = 0$, $\tau_0 = 0.3$, $k = \tau_0/H$ km$^{-1}$, gives $D \ll {\sim}17$ km. For this range of $D$, then, Eq. (A3.9) becomes:

$$L(\mathbf{0}; D) = C_0 \, A_s \left(\frac{1}{D}\right) \int_{x=x_d}^{x_H} \frac{1}{1+x^2} \, dx = C_0 \, A_s \left(\frac{1}{D}\right) \left[\tan^{-1}\left(\frac{H}{D}\right) - \tan^{-1}\left(\frac{h_d}{D}\right)\right] \quad (A3.12)$$

It is then expected that close to the source, starting approximately from the distance for which $D^2 + h_d{}^2 > A_s$, and up to the first few hundred meters away from it ($D \sim 0.1$ km, $H/D \sim 80$, $\tan^{-1}(H/D) \approx \pi/2$ to within a 1%, $h_d/D \leq 0.1$, $\tan^{-1}(h_d/D) \approx h_d/D$ to within a 1%) the artificial zenith brightness will decrease with $D$ and $h_d$ as:

$$L(\mathbf{0}; D) \approx C_0 \, A_s \left(\frac{1}{D}\right)\left[\frac{\pi}{2} - \frac{h_d}{D}\right] \quad (A3.13)$$

which, in logarithmic scale is equivalent to

$$\log_{10} L(\mathbf{0}; D) \approx \log_{10}(C_0 \, A_s) - \log_{10}(D) + \log_{10}\left(\frac{\pi}{2} - \frac{h_d}{D}\right) \quad (A3.14)$$

which, for an observer at the same altitude as the source ($h_d = 0$) becomes a simple linear dependence on $\log_{10}(D)$ with slope -1, and for $h_d > 0$ gives

$$\log_{10} L(\mathbf{0}; D) \approx \log_{10}(C_0 \, A_s) - \log_{10}(D) + \log_{10}\left(1 - \frac{2h_d}{\pi D}\right) + \log_{10}\frac{\pi}{2} \quad (A3.15)$$

which, for a fixed value of $D$ decreases linearly with $h_d$ for small values of the ratio $h_d/D$, with slope $-2/[\pi D \log_e(10)]$.

For a more detailed account of the asymptotic behavior of the artificial scattered radiance, including the $1/D$ dependence, the reader is referred to the comprehensive treatment developed by Kocifaj et al [41].

**Author contributions**

**SB:** Conceptualization, Methodology, Numerical calculations, Formal analysis, Writing - Original Draft, Writing - Review & Editing; **CB-V:** Conceptualization, Methodology, Formal analysis, Writing - Review & Editing; **MK:** Conceptualization, Methodology, Formal analysis, Writing - Review & Editing

**Funding**

CB-V acknowledges funding from Xunta de Galicia/FEDER, grant ED431B 2020/29. MK was supported by the Slovak Research and Development Agency (grant number APVV-18-0014) and the Slovak National Grant Agency VEGA (grant no. 2/0010/20).

[9] Kyba CCM, Garz S, Kuechly H, Sánchez de Miguel A, Zamorano J, Fischer J, Hölker F. High-Resolution Imagery of Earth at Night: New Sources, Opportunities and Challenges. Remote Sens 2015;7:1-23. doi: 10.3390/rs70100001

[10] Levin N, Kyba CCM, Zhang Q, Sánchez de Miguel A, Román MO, Li X, Portnov BA, Molthan AL, Jechow A, Miller SD, Wang Z, Shrestha RM, Elvidge CD. Remote sensing of night lights: A review and an outlook for the future. Remote Sens Environ 2020;237:111443. doi: 10.1016/j.rse.2019.111443

[11] Li X, Elvidge C, Zhou Y, Cao C, Warner T. Remote sensing of night-time light. Int J Remote Sens 2017;38(21):5855-5859. doi: 10.1080/01431161.2017.1351784

[12] Román MO, Wang Z, Sun Q, Kalb V, Miller SD, Molthan A, Schultz L, Bell J, et al. NASA's Black Marble nighttime lights product suite. Remote Sens Environ 2018;210:113-143. doi: 10.1016/j.rse.2018.03.017

[13] Rybnikova N, Sánchez de Miguel A, Rybnikov S, Brook A. A New Approach to Identify On-Ground Lamp Types from Night-Time ISS Images. Remote Sens 2021;13(21):4413. doi: 10.3390/rs13214413

[14] Sánchez de Miguel A, Kyba CCM, Aubé M, Zamorano J, Cardiel N, Tapia C, Bennie J, Gaston KJ. Colour remote sensing of the impact of artificial light at night (I): The potential of the International Space Station and other DSLR-based platforms. Remote Sens Environ 2019;224:92–103. doi: 10.1016/j.rse.2019.01.035

[15] Stefanov WL, Evans CA, Runco SK, Wilkinson MJ, Higgins MD, Willis K. 2017. Astronaut Photography: Handheld Camera Imagery from Low Earth Orbit, in J.N. Pelton et al. (eds.), Handbook of Satellite Applications, Springer International Publishing Switzerland, doi: 10.1007/978-3-319-23386-4_39

[16] Zhu X, Tan X, Liao M, Liu T, Su M, Zhao S, Xu YN, Liu X. Assessment of A New Fine-Resolution Nighttime Light Imagery from the Yangwang-1 ("Look Up 1") Satellite. IEEE Geosci Remote Sens Lett 2022;19:1-5. doi: 10.1109/LGRS.2021.3139774

[17] Simoneau A, Aubé M, Leblanc J, Boucher R, Roby J, Lacharité F. Point spread functions for mapping artificial night sky luminance over large territories. Mon Notices Royal Astron Soc 2021;504(1):951–963. doi: 10.1093/mnras/stab681
32